\documentclass[sigconf]{acmart}

\usepackage{amsmath,amsfonts,bm}

\def\eqref#1{equation~\ref{#1}}

\def\1{\bm{1}}

\DeclareMathAlphabet{\mathsfit}{\encodingdefault}{\sfdefault}{m}{sl}
\SetMathAlphabet{\mathsfit}{bold}{\encodingdefault}{\sfdefault}{bx}{n}

\newcommand{\sigmoid}{\sigma}

\usepackage[ruled,vlined]{algorithm2e}
\usepackage{pdfpages}
\usepackage{import}
\usepackage{paralist}
\usepackage{subcaption}
\usepackage{caption}
\usepackage{multirow}
\usepackage{makecell}
\usepackage{threeparttable}

\AtBeginDocument{%
  \providecommand\BibTeX{{%
    \normalfont B\kern-0.5em{\scshape i\kern-0.25em b}\kern-0.8em\TeX}}}

\setcopyright{acmcopyright}
\copyrightyear{2018}
\acmYear{2018}
\acmDOI{10.1145/1122445.1122456}

\acmConference[XXX]{XXXX}{June 03--05, 2021}{XXXX}
\acmBooktitle{WXXXX,
  June 03--05, 2021, XXX, XX}
\acmPrice{15.00}
\acmISBN{978-1-4503-XXXX-X/18/06}

\graphicspath{./img}

\begin{document}


\title{Multi-behavior Graph Contextual Aware Network for Session-based Recommendation}

\author{Qi Shen}
\authornote{Both authors contributed equally to this research.}
\affiliation{%
  \institution{Tongji University}
  \city{Shanghai}
  \country{China}}
\email{1653282@tongji.edu.cn}

\author{Lingfei Wu}
\authornotemark[1]
\affiliation{%
  \institution{JD Silicon Valley Research Center}
  \country{United States}}
\email{lwu@email.wm.edu}

\author{Yitong Pang}
\affiliation{%
  \institution{Tongji University}
  \city{Shanghai}
  \country{China}}
\email{1930796@tongji.edu.cn}

\author{Yiming Zhang}
\affiliation{%
  \institution{Tongji University}
  \city{Shanghai}
  \country{China}}
\email{2030796@tongji.edu.cn}

\author{Zhihua Wei}
\authornote{Corresponding author.}
\affiliation{%
  \institution{Tongji University}
  \city{Shanghai}
  \country{China}}
\email{zhihua_wei@tongji.edu.cn}

\author{Fangli Xu}
\affiliation{%
  \institution{Squirrel AI Learning}
  \country{United States}}
\email{fxu02@email.wm.edu}

\author{Bo Long}
\affiliation{
  \institution{JD.COM}
  \country{United States}}
\email{bo.long@jd.com}

\renewcommand{\shortauthors}{Zhang, et al.}

\begin{abstract}
Predicting the next interaction of a short-term sequence is a challenging task in session-based recommendation (SBR).
Multi-behavior session recommendation considers session sequence with multiple interaction types, such as click and purchase, to capture more effective user intention representation sufficiently.
Despite the superior performance of existing multi-behavior based methods for SBR, there are still several severe limitations:
(i)  Almost all existing works concentrate on single target type of next behavior and fail to model multiplex behavior sessions uniformly.
(ii) Previous methods also ignore the semantic relations between various next behavior and historical behavior sequence, which are significant signals to obtain current latent intention for SBR.
(iii) The global cross-session item-item graph established by some existing models may incorporate semantics and context level noise for multi-behavior session-based recommendation.

To overcome the limitations (i) and (ii), we propose two novel tasks for SBR, which require the incorporation of both historical behaviors and next behaviors into unified multi-behavior recommendation modeling.
To this end, we design a Multi-behavior Graph Contextual Aware Network (MGCNet) for multi-behavior session-based recommendation for the two proposed tasks. 
Specifically, we build a multi-behavior global item transition graph based on all sessions involving all interaction types.
Based on the global graph, MGCNet attaches the global interest representation to final item representation based on local contextual intention to address the limitation (iii).
In the end, we utilize the next behavior information explicitly to guide the learning of general interest and current intention for SBR. 
Experiments on three public benchmark datasets show that MGCNet can outperform state-of-the-art models for multi-behavior session-based recommendation.
\end{abstract}




\maketitle

\section{Introduction}

Recommendation systems are widely used in online platforms, as an infrastructural tool to select interesting information for users and alleviate information overload. Recently, in feed streaming applications including media streaming (e.g., TikTok and Spotify) and e-commerce (e.g., Amazon and Alibaba), recommendation systems need to focus on the interactions within the current active session for satisfactory recommendation results. Conventional recommendation methods that usually learn static user preferences from the long-term  historical interactions, e.g., collaborative filtering \cite{sarwar2001item}, are typically not suitable for these scenarios. Therefore, session-based recommendation (SBR) where the recommendation relies solely on the user’s behavioral sequence in the current session, has attracted great attention in the past few years. Generally, SBR aims to predict next item a user would likely to consume, based on the recent items interacted by the user in an ongoing session. Inspired by the advancement of deep learning techniques, various neural network-based models have been developed to capture the short-term intention of users, and have achieved promising results \cite{gru4rec,srgnn}.


Despite the prevalence of the above SBR methods, most of them concentrate on capturing user preference based on single type of interactive behavior sequence solely.
However, in practical recommendation tasks, there are naturally multiple behavior interactions in user sessions with complex behavior transition patterns, which are overlooked by the common SBR methods.

Take the e-commerce system as an example, in an ongoing session, a user may click, collect or even purchase goods according to the current personal intention. 
Intuitively, different behavior types reflect different user intention strengths.
For instance, the adding-to-favorite behavior is a stronger intention signal compared to the click behavior.
Moreover, various behavior patterns may reveal implicit item relationships. 
For example, the item transitions with \emph{click2click} or \emph{click2purchase} behavior pattern may be replaceable (e.g., click \emph{iPhone} then click \emph{SAMSUNG phone}), while the item transitions with \emph{purchase2purchase} or \emph{purchase2click} pattern may be complementary (e.g., click item \emph{iPhone} then purchase \emph{MacBook}). Therefore, it is a challenging but valuable work to effectively model multi-behavior sequence for SBR.

\begin{figure}[t]
    \centering
    \includegraphics[width=0.45\textwidth]{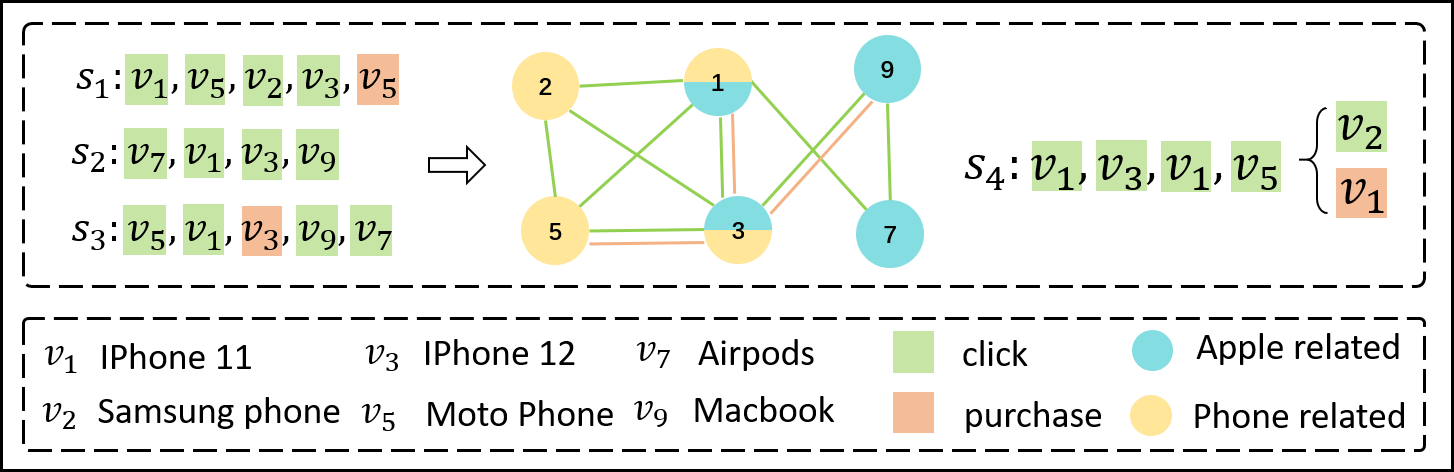}
    \caption{A toy example of Multi-Behavior SBR.}
    \label{fig:exmaple}
    \vspace{-0.5cm}
\end{figure}

Despite the existence of a handful of models that integrate multi-behavior information for SBR \cite{beyondclick, MKMSR}, these works suffer from two limitations:
\begin{itemize}
    \item \textbf{The recommendation is limited to the specified target behavior. } Traditional frameworks for multi-behavior session-based recommendation (MBSBR) usually allocate a specific interaction type as target behavior and explore other behavior data as auxiliary information. This restriction of the target behavior contributes to deficiency of generalization to other target behavior session recommendations and fails to achieve the unified modeling for multiplex behavior for next user-item interaction. Meanwhile, the sparsity of the target behavior data may limit the capability of learning generally item transitions comprehensively. More importantly, the session recommendation with presupposed target behavior deviates from the real scene, which is unknown for next behavior type. The designation of target behavior seriously limits the application of these methods.
    \item \textbf{ The relation of next behavior and historical behaviors are not comprehensively explored.} Existing methods encode the current multi-behavior relational session without the injection of next behavior, which ignore the intention strength signals for next item and contextual semantics information between next behavior and former behavior sequence. For instance,  as illustrate in Figure \ref{fig:exmaple}, different next behavior types will lead to different next interaction items of session $s_4$. For current user who has explored phone-related items, the next click behavior indicates the further exploration intention with other uninteracted items like $v_2$, while next purchase behavior implies the summative exploitation intention which prefers to the historical item in current session, like $v_1$.
\end{itemize}



Meanwhile, many works constructed a global item-item graph, consists of item-transitions overall sessions, and utilized variations of GNNs to enhance the item representations by augmenting cross-session interaction patterns for SBR \cite{gce-gnn,beyondclick,oderunorder}.
For multi-behavior session data, multi-relational behavior pattern is additionally introduced to global item graph as edge attributes.
However, the injection of other item-transitions information will incorporate two levels of irrelevant noise for current session intention representation learning:
(i)
\emph{semantics level}, which may aggregate other irrelevant behavior pattern information for current item representation under the related  behavior modality. Especially, diverse behavior patterns with different item relations make different contributions to current behavior.
(ii)
\emph{context level}, which may introduce irrelevant contextual information to current session representations caused by the information loss of the conversion from different sessions to pairwise item transitions without distinctive session information.
We illustrate this with an example in Figure \ref{fig:exmaple}.
Generally, after the encoding of GNN which ignores the local transition of current session, the global item representation of $v_3$ contains various item transition information with different interest, i.e.,  \emph{Apple-related} interest and  \emph{phone-related} interest, and different behavior patterns semantics, i.e.,  \emph{purchase2purchase} and  \emph{click2click}, which conflicts with the original phone-related intention of clicking item $v_3$ within current session $s_4$.

Motivated by the aforementioned challenges, this work proposes two general tasks for MBSBR.
The first task offers all kinds of next behavior types for item recommendation in a unified way, which extends the single target behavior based task of mainstream researches.
More generally, the second task masks the next behavior type consistent with real scenes, and requires an explicit prediction of next behavior through historical interaction data to guide the prediction of next item.
To address the above tasks, we propose a Multi-behavior Graph Contextual aware Network for MBSBR.
Following previous works for SBR, we construct a multi-behavior global item transition graph based on all sessions involving all interaction types.
Based on the global graph, our model attaches the global interest representation to final item representation based on local contextual intention.
In the end, we utilize the next behavior information explicitly to guide the learning of general interest and current intention for SBR. 


The contributions of this work are summarized as follows:
\begin{itemize}
    \item We propose two novel tasks for multi-behavior session-based recommendation, which require the incorporation of both historical behaviors and next behaviors into unified multi-behavior recommendation modeling.
    \item To address two proposed tasks, we develop a unified Multi-behavior Graph Contextual aware Network (MGCNet), which integrates global intentions of global transitions overall sessions and local intentions of current session, and injects the significant next behavior signals into session representation explicitly.
    \item Extensive experiments on three datasets demonstrate that our model is superior compared with state-of-the-art models for MBSBR.
\end{itemize}


\begin{figure*}
    \centering
    \includegraphics[width=0.75\textwidth]{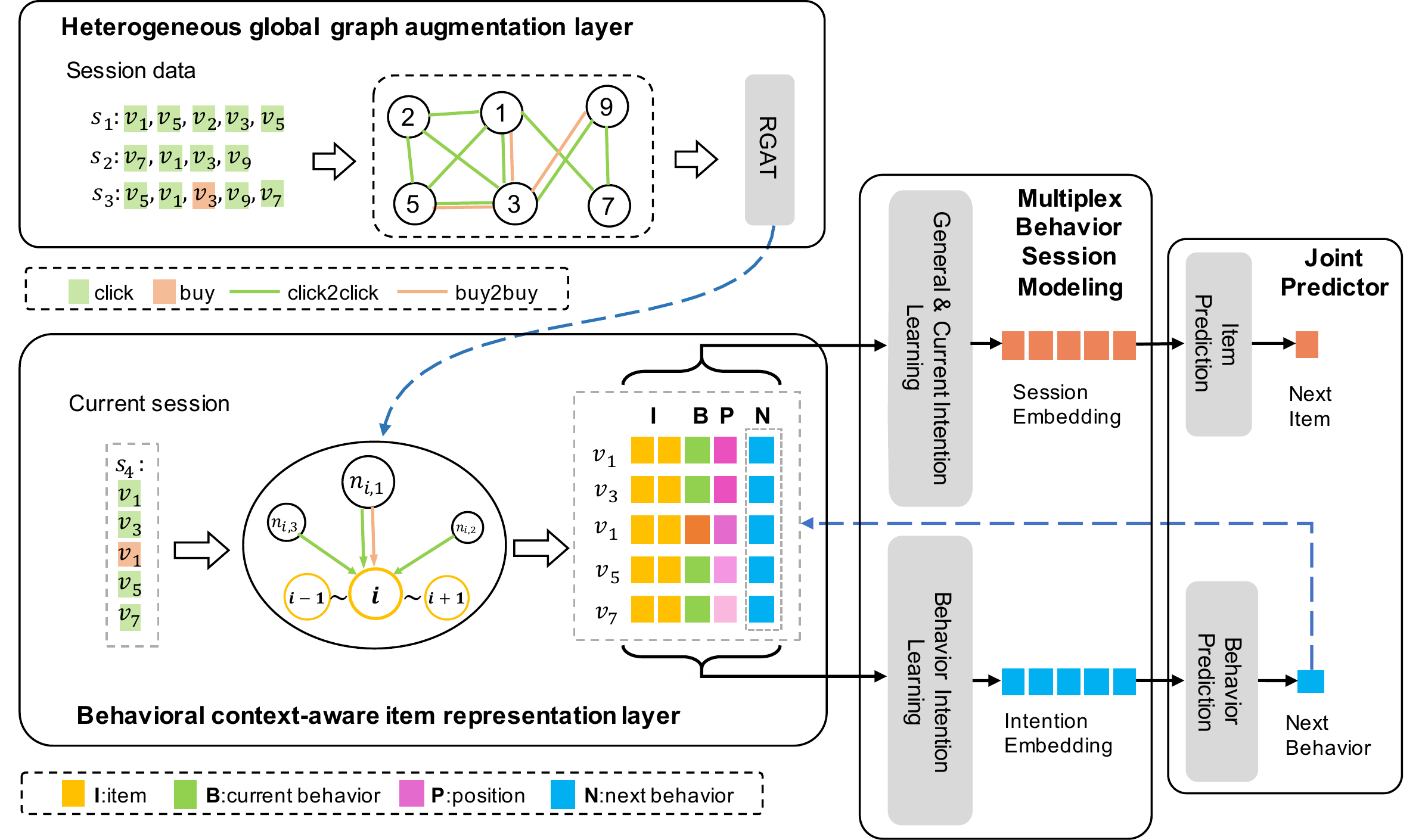}

    \caption{The overview of MGCNet.}
    \label{fig:framework}
    \vspace{-0.45cm}
\end{figure*}

\section{Related Work}

\textbf{Session-based Recommendation.}
Early SBR methods widely used Markov chains to capture the sequential signal in session sequences \cite{rendle2010factorizing,wang2015learning}. Following the development of deep learning, many neural network based  approaches have been proposed for session-based recommendation. Due to the sequence modeling capability of recurrent neural networks (RNNs),  RNN-based methods have been the most popular methods for SBR \cite{gru4rec,narm,liu2018stamp,2020s3,ren2019repeatnet,song2019islf}. For instance, GRU4Rec \cite{gru4rec} was firstly proposed to utilize GRU layer to capture item interaction sequences. 
Based on GRU4Rec, NARM \cite{narm} utilized the attention mechanism based on last interaction item after RNN to capture the global and local  preference representation of the user in the current session. 

More recently, a large number of methods relying on graph structure have appeared for recommendation. Motivated by the powerful capability to extract complex relationships between objects of graph neural networks (GNNs),  these methods proposed to utilize the GNNs to extract the item transition patterns for SBR \cite{srgnn,xu2019graph,fgnn,stargnn,lessr,gce-gnn,pang2021heterogeneous}. The first work is SR-GNN \cite{srgnn}, which converted the interaction sequence into a directed graph and employed the gated GNN (GGNN) on the session graph to learn item embedding. GC-SAN \cite{xu2019graph} further  adopted self-attention mechanism to capture global dependencies among different positions and integrated GGNN to generate the session embedding. To form better graph structure from the session, LESSR \cite{lessr} proposed a lossless encoding scheme which preserved the edge-order and sequential information. GCE-GNN \cite{gce-gnn} exploited two GNN modules to learn two level item representations separately: the session-level item embedding based on current session graph and the global-level item embedding based on global item-transition graph, and then integrated different-level item embeddings  into the final session representation based on the position-aware attention.

\textbf{Multi-behavior Recommendation.} Conceptually, multi behavior recommendation models multiple types of user-item feedback for enhancing recommendation on target behaviors \cite{MBMT,MBGCN,chen2020efficient,tang2016empirical,gu2020hierarchical,zhou2018micro,gao2019neural}.
Current mainstream research employed the collaborative filtering or GNN-based models to capture the user-item interaction feature from multiplex behaviors. Gao et al. \cite{MBMT} proposed a tailored collaborative filtering model which correlated the multi-behavior prediction in a cascaded way. As a GNN-based method, MBGCN \cite{MBGCN} constructed a heterogeneous user-item graph to represent multi-behavior interactions, and captured the item-to-item similarity and user-to-item relations to model the semantics and influence of multiplex behavior. Besides, some works focus on modeling the interaction order with multiplex behavior information for sequence recommendation. For instance, KHGT \cite{MBHGT} proposed to distill the type-specific user-item interactive patterns by graph-structured transformer module at first, then encoded the cross-type behavior hierarchical dependencies by an attentive fusion network.

Distinguished from above SBR methods which capturing  current user preference based on singular type  of interactive behavior sequence, multi-behavior session-based recommendations predict the next item based on the multiplex behavior session sequence. Currently, multi-behavior session-based recommendation is still in the early stage of research. MGNN-SPred \cite{beyondclick} firstly  proposed to learn global item-to-item relations through GNN and integrate embedding of target and auxiliary of current session by the gating mechanism. MKM-SR \cite{MKMSR} incorporated user micro-behaviors into session modeling to capture the transition pattern, modeled the item relation and the behavior sequence through the GGNN and GRU.

However, these methods usually regarded historical auxiliary user-item interactions as weak signals for forecasting next item with target behavior, which merely utilized all behavioral data for joint learning through a unified model without limitation to designated next target behavior.
Besides, the complicated relations between next behavior and historical behaviors have not been comprehensively considered.

\section{Preliminary}
In this work, we aim to explore the abundant semantic relations among multiplex user behaviors in current session. 

Given the entire session set $\mathcal{S}$, item set $\mathcal{V}$ and interaction behavior set $\mathcal{O}$, we first define a multi-behavior item sequence as $s=\{(v_1,o_1),(v_2,o_2),...,(v_{|s|},o_{|s|})\in S$, where $(v_i,o_i)$ represents the user interacted item $v_i \in \mathcal{V}$ with the behavior $o_i \in \mathcal{O}$.

\textbf{Problem Statement.} To achieve the unified modeling for diverse next behavior SBR, we present the formulation of the problems researched in this paper as below.
\begin{itemize}
\item \textbf{Task 1: Next-item Prediction with Given next Behavior.} Given a multi-behavior session $s$, the expectation of this task is to predict the next item $v_{|s|+1}$ with provided next interaction type $o_{|s|+1} \in \mathcal{O}$.
In other words, given a session with target next behavior $o_{|s|+1}$, here task 1 for MBSBR is to predict the next item $v_{|s|+1}$ that the user is most likely to interact with the provided target behavior $o_{|s|+1}$. This task extends the common  single target behavior based task of mainstream MBSBR researches in a unified way for diverse next behaviors.

\item \textbf{Task 2: General Next-item Prediction.} Given a session $s \in S $ without the restrictions of specific target behavior, we explore to predict the next user-item interaction type $o_{|s|+1} \in \mathcal{O}$, moreover, estimate the probability of next item $v_{|s|+1}$ under the predicted behavior $o_{|s|+1}$. In other words, the model not only needs to predict the next behavior (e.g., buy or click) but also give should predict the next item under the predicted next behavior (e.g., next item the user will click or buy).
\end{itemize}


\section{Methodology}
\subsection{Overview}
As shown in \autoref{fig:framework}, MGCNet contains four main components: Heterogeneous global graph augmentation layer, Behavioral context-aware item representation layer, Multiplex behavior session modeling layer, and joint predictor. 
With the global graph augmentation and context-aware item representation layer, each item representation integrates the global transition patterns of all users with activated contextual information.
Furthermore, at the multiplex behavior session modeling layer, we obtain the user’s latent interest and behavior motivations respectively. 
With the injection of given next behavior from data or feed by joint predictor, our model not only considers the historical behavior sequence but also the next latent behavior.
Finally, the joint predictor estimates the probability of candidate items and the interaction types.
In particular, our model is mainly designed for Task 2, but it can also be applied to the basic next-item prediction with target behavior for Task 1.

\subsection{Heterogeneous Global Graph Construction}
As mentioned above, the aggregation of diverse item transitions over all sessions may augment the item representations for SBR.
And the abundant multi-behavior relations are conducive to capture the latent intentions.
Therefore, we construct the global heterogeneous multi-behavior item transition graph $\mathcal{G}=(\mathcal{V},\mathcal{E})$ by taking all items as nodes $\mathcal{V}$ and each type of cross-behavior patterns as edges $\mathcal{E}$ from training session sequences.
For instance, the sequential operation of clicking item $v_i$ and buying item $v_{i+1}$ in one session is converted to the adjacent relation $(v_{i},click2buy,v_{i+1})$ in $\mathcal{G}$. 
Furthermore, for capturing long-distance item transitions with same behavior, we introduce the same interaction based adjacent dependency
$\{(v_{i},o_i2o_j,v_{j}) \, |\, o_i=o_j, |i-j|>1,(v_{i},o_i),(v_{j},o_j) \in s_k,s_k\in S\}$ to $\mathcal{G}$ explicitly.

\subsection{Heterogeneous Global Graph Augmentation}
We next present how to propagate features on heterogeneous global graph to encode item-transition information from all behavioral sessions to improve the accuracy of recommendation.
Our Graph Encoder firstly aggregates the neighbor nodes with each relation respectively and gathers the inner-semantic information as updated node representations for heterogeneous multi-relation graph following \cite{rgcn}. 
Let $\mathbf{h}_{i}^{(k)}$ denote the embedding of item $v_i$ after $k$ layers GNN propagation.
The item IDs are embedded in $d$-dimensional space and are used as initial node features in our model, $\mathbf{h}_{i}^{(0)}\in \mathbb{R}^d$.
Following \cite{gat}, we calculate the attention weights based on the similarity of each item transition and the statistic times of occurrences for each relational subgraph, respectively. The aggregation process of each relation can be denoted as follows:
\begin{equation}
\setlength{\abovedisplayskip}{3pt}
\setlength{\belowdisplayskip}{3pt}
\begin{aligned}
    e_{ij}^{r,(k)}=\mathbf{a}^{(k)}_r\left( \mathbf{W}^{(k)}_{r,t} \mathbf{h}_{i}^{(k)}+\mathbf{W}^{(k)}_{r,s} \mathbf{h}_{j}^{(k)}+ \mathbf{W}^{(k)}_{r,w} w^r_{ij} \right)\,,
\end{aligned}\label{eqn:grad}
\end{equation}
\begin{equation}
\setlength{\abovedisplayskip}{3pt}
\setlength{\belowdisplayskip}{3pt}
\begin{aligned}
    \alpha_{ij}^{r,(k)}=\underset{j\in \mathcal{N}(i)}{\text{softmax}}\left(\mathrm{LeakyReLU}\left(e_{ij}^{r,(k)}\right)\right)\,,
\end{aligned}\label{eqn:grad}
\end{equation}
\begin{equation}
\setlength{\abovedisplayskip}{3pt}
\setlength{\belowdisplayskip}{3pt}
\begin{aligned}
    \mathbf{h}_{r,i}^{(k+1)}=\sum_{j\in \mathcal{N}_r (i)}{\alpha_{i,j}^{r,(k)}}\mathbf{W}_{r,s}^{(k)} \mathbf{h}_{j}^{(k)}\,,
\end{aligned}\label{eqn:grad}
\end{equation}
$r\in\mathcal{R}$ is the specific relation for current subgraph, e.g. \emph{click2click}, $w^r_{ij}$ is the weight of edge $(v_i,r,v_j)$ in global graph, $\mathbf{W}^{(k)}_{r,s}, \mathbf{W}^{(k)}_{r,t}\in \mathbb{R}^{d\times d} $ and $\mathbf{a}^{(k)}_r, \mathbf{W}^{(k)}_{r,w} \in \mathbb{R}^{d} $ are trainable parameters.

After attention-based inner-relation propagation model, we accumulate all behavior pattern messages to generate cross-relation representation. 
Meanwhile, we add self-connections to maintain the information mobility with corresponding representation at previous layer.
\begin{equation}
\setlength{\abovedisplayskip}{3pt}
\setlength{\belowdisplayskip}{3pt}
\begin{aligned}
    \mathbf{h}_{i}^{(k+1)}=\frac{1}{|\mathcal{R}|}\sum_{r\in\mathcal{R}}{\mathbf{h}_{r,i}^{\left( k+1 \right)}}+\mathbf{W}_{res}^{\left( k \right)}\mathbf{h}_{i}^{\left( k \right)}\,,
\end{aligned}\label{eqn:grad}
\end{equation}
 where $R$ indicates the set of all behavior patterns, $\mathbf{W}_{res}^{\left(k\right)}$ maps the previous layer embedding to current semantic space. 
After the message passing of inner-semantic and cross-semantic on heterogeneous multi-behavior global graph, we obtain the global representation ${\mathbf{h}_i^g}={\mathbf{h}_i^{(k+1)}}$, which is dependent on itself and its immediate adjacent items with multiplex behavior patterns.

\subsection{Behavioral Context-Aware Item Representation Learning}
The item representation learnt from heterogeneous global graph overlooks the contextual information in current multi-behavior session. Therefore,  we propose a method called Behavioral Context-Aware Attention (BCAN), which integrates rich global semantic information with specific session related item embeddings. The core idea is to estimate the relevance between neighbor items and user's local intentions based on the item transitions and behavior patterns.

To capture the local user intentions in current session, we employ a GRU model to encode the interaction item sequence with multi-behavior information:
\begin{equation}
\setlength{\abovedisplayskip}{3pt}
\setlength{\belowdisplayskip}{3pt}
\begin{aligned} 
    \left(\mathbf{c}_{v_1},\mathbf{c}_{v_2},...,\mathbf{c}_{v_t}\right)=\text{GRU}\left(\mathbf{q}_{v_1^s}^{},\mathbf{q}_{v_2^s}^{},...,\mathbf{q}_{v_L^s}^{}\right)\,,
\end{aligned}\label{eqn:grad}
\end{equation}
where $\mathbf{q}_{v_i^s}=\left[\mathbf{h}_{v_i^s}^{(0)} \parallel \mathbf{o}_{v_i^s}\right]$ concatenates the initial item embedding $\mathbf{h}_{v_i^s}^{(0)}$ with behavior type embedding $\mathbf{o}_{v_i^s}$.

Based on the representation of the local interest, we firstly calculate the relevance score $\alpha_{ij}$ between the local intention of item $i$ and the general attraction of item $j$:
\begin{equation}
\setlength{\abovedisplayskip}{3pt}
\setlength{\belowdisplayskip}{3pt}
\begin{aligned} 
    \alpha_{ij}=\underset{j\in \mathcal{N} (i)}{\text{softmax}}\left(\text{LeakyReLU}\left(\text{cos}\left(\mathbf{W}_l \mathbf{c}_{v_i},\mathbf{W}_g \mathbf{h}^g_{j}\right)\right)\right)\,,
\end{aligned}\label{eqn:grad}
\end{equation}

Then we aggregate neighbor item embeddings together based on weight $\alpha_{ij}$ to obtain the final behavioral context augmented item representations on the current session. 
Finally, we combine the local and global item representations by hyper-parameter intention factor $\beta$ as follows:
\begin{equation}
\begin{aligned} 
    \mathbf{h}_{v^s_i}^{l}=\sum_{j \in \mathcal{N}(i)}{\alpha_{ij}\mathbf{h}_j^g} \,,
\end{aligned}\label{eqn:grad}
\end{equation}
\begin{equation}
\begin{aligned} 
         \mathbf{h}_{v^s_i}^{}=\mathbf{h}_{v^s_i}^{g}+\beta \mathbf{h}_{v^s_i}^{l}\,,
\end{aligned}\label{eqn:grad}
\end{equation}

Through the module above, we obtain behavioral contextual-aware item representations with the  reduction of unrelated global intentions and the enhancement of corresponding interaction information.

\subsection{Multiplex Behavior Session Modeling}

To obtain a session representation with multiple user feedback types, the fusion of behavior information and item transitions in current session requires comprehensive consideration.
As discussed above, it is essential to model the behavior-specific intention strength and cross-behavior contextual information for MBSBR, involving the next behavior tag and historical interaction types.
Meanwhile, the next intention of user is dynamic and not necessarily continuous with last item.
Therefore, it is reasonable to portray session representation with the historical intention and the current intention.
To meet the above requirements synchronously, we propose this module to capture the flexible item transitions and behavior patterns.

To learn the historical interest of the current session, we employ a general interest based attention mechanism. In detail, given the item $v_i$ in the current session, we concatenate the item representation $\mathbf{h}_{v_i}$ learnt from the BCAN module with the embeddings of the corresponding behavior, the given target behavior type and the reversed position:
\begin{equation}
\setlength{\abovedisplayskip}{3pt}
\setlength{\belowdisplayskip}{3pt}
\begin{aligned}
    \mathbf{m}_{v_i^s}=\text{tanh}(\mathbf{W}_0(\mathbf{h}_{v_i^s}\parallel \mathbf{o}_{v_i^s} \parallel \mathbf{o}_{v_{L+1}^s} \parallel \mathbf{p}_{L-i+1}))\,,
\end{aligned}\label{eqn:graph}
\end{equation}
where behavior type embedding $\mathbf{o}_{v_i^s} \in \mathbb{R}^d$ and reverse position embedding $\mathbf{p}_{L-i+1} \in \mathbb{R}^d$ are trainable initial parameters.
Here, reversed position \cite{gce-gnn} reflects the chronological order of item transitions from last to first.
Moreover, a pseudo target behavior tag is predicted in Task 2, which depends on prediction module and will be represented in next part.
Then, the correlation of each behavioral item and general historical intentions is obtained by the  attention mechanism based on coarse-level average pooling 
$\mathbf{s}'=\frac{1}{L}\sum_{i=1}^L\mathbf{m}_{v_i^s}$ of all items in this session: 
\begin{equation}
\setlength{\abovedisplayskip}{3pt}
\setlength{\belowdisplayskip}{3pt}
\begin{aligned}
    \beta_i=\mathbf{r}^T\sigmoid(\mathbf{W}_1\mathbf{m}_{v_i^s}+\mathbf{W}_2\mathbf{s}'+\mathbf{b}_{g})\,,
\end{aligned}\label{eqn:attt}
\end{equation}
\begin{equation}
\setlength{\abovedisplayskip}{3pt}
\setlength{\belowdisplayskip}{3pt}
\begin{aligned}    
    \mathbf{s}_{g}=\sum_{i=1}^L\beta_i\mathbf{m}_{v_i^s} \,,
\end{aligned}\label{eqn:attt}
\end{equation}
where $\sigmoid$ is an activate function, $\mathbf{W}_1\,, \mathbf{W}_2 \in \mathbb{R}^{d \times d}$ and $\mathbf{r}\,,\mathbf{b}_{g}\in \mathbb{R}^d$ are trainable parameters. 

To simplify the formula, we use $\text{Attention}\left(\left\{ \mathbf{m}_{v_i^s}|i=1,\dots,L\right\},\mathbf{s}'\right)$ to denote the attention module, which calculates the attention scores between each item $\mathbf{m}_{v_i^s}$ and anchor embedding $\mathbf{s}'$ like formula \autoref{eqn:attt}.

Analogously, the current intention of user is learnt by an item-level soft-attention mechanism:
\begin{equation}
\setlength{\abovedisplayskip}{3pt}
\setlength{\belowdisplayskip}{3pt}
\begin{aligned}
     \mathbf{s}_{c}=\text{Attention}\left(\left\{ \mathbf{m}_{v_i^s}|i=1,\dots,L\right\},\mathbf{m}_{v_L^s}\right)\,,
\end{aligned}\label{eqn:graph}
\end{equation}

Then we combine the current intention and the historical preference representation to generate the final session representation:
\begin{equation}
\setlength{\abovedisplayskip}{3pt}
\setlength{\belowdisplayskip}{3pt}
\begin{aligned}
    \mathbf{S}=\sigmoid(\mathbf{W}_c(\mathbf{s}_{g} \parallel \mathbf{s}_{c}^s)\,
\end{aligned}\label{eqn:graph}
\end{equation}

To infer the next behavioral intention, we also apply a local intention anchored soft-attention mechanism as follows:
\begin{equation}
\begin{aligned}
    \mathbf{C}=\text{Attention}\left(\left\{\mathbf{c}_{v_i^s}|i=1,\dots,L\right\},\mathbf{c}_{v_L^s}\right)\,,
\end{aligned}\label{eqn:graph}
\end{equation}
\begin{equation}
\setlength{\abovedisplayskip}{3pt}
\setlength{\belowdisplayskip}{3pt}
\begin{aligned}
    \mathbf{B}=\sigmoid\left( \mathbf{W}_{bhv} \mathbf{C}+\mathbf{b}_{bhv} \right)\,,
\end{aligned}\label{eqn:graph}
\end{equation}
where $\mathbf{W}_{bhv}$ and $\mathbf{b_{bhv}}$ are trainable parameters.



\subsection{Predictor and Training}\label{sec:predictor}
Intuitively, the calibration of next behavior can help capture the next interaction intention more effectively. For Task 1, the target behavior tag is offered following the definition, and all we need to do is estimate the probability of candidate items forced by the ground truth behavior type. 
For Task 2, we firstly predict the next behavior based on the historical behavior patterns and item transitions, then feed it to session representation module as a pseudo-tag to assist the next-item prediction.
To predict the next interaction type, we estimate the probability of all kind of behavior based on current behavioral intention representation:
\begin{equation}
\setlength{\abovedisplayskip}{3pt}
\setlength{\belowdisplayskip}{3pt}
\begin{aligned}
    \hat{{b}}_i=\text{softmax}(\mathbf{B}^T \mathbf{o}_{i})\,,
\end{aligned}\label{eqn:graph}
\end{equation}
Then we select the max probability behavior type as the presumed next behavior $o_{v_{L+1}^s}=\max\limits_{i} \hat{{b}}_i $ for Task 2.

Based on the session embedding $\mathbf{S}$ obtained above and the initial embeddings of candidate items, we can compute the recommendation probability $\hat{\mathbf{y}}$ of candidate items:
\begin{equation}
\setlength{\abovedisplayskip}{3pt}
\setlength{\belowdisplayskip}{3pt}
\begin{aligned}
    \hat{{y}}_i=\text{softmax}(\mathbf{S}^T\mathbf{h}_{i}^{(0)})\,,
\end{aligned}\label{eqn:graph}
\end{equation}
where $\hat{{y}}_i \in \hat{\mathbf{y}}$ denotes the probability that the user will interact  with item $v_i \in V$ in the current session. 

The objective function can be formulated as a cross entropy loss as follows, which consists of next item loss $\mathcal{L}_{item}$ and next behavior loss $\mathcal{L}_{bhv}$:
\begin{equation}
\setlength{\abovedisplayskip}{3pt}
\setlength{\belowdisplayskip}{3pt}
\begin{aligned}                                     
    \mathcal{L}_{item}=-\sum_{i=1}^{|V|}{y}_i\text{log}(\hat{{y}}_i)+(1-{y}_i)\text{log}(1-\hat{{y}}_i)\,,
\end{aligned}
\end{equation}        
\begin{equation}
\setlength{\abovedisplayskip}{3pt}
\setlength{\belowdisplayskip}{3pt}
\begin{aligned}                     
    \mathcal{L}_{bhv}=-\sum_{i=1}^{|R|}\lambda_i{b}_i\text{log}(\hat{{b}}_i)+(1-{b}_i)\text{log}(1-\hat{{b}}_i)\,,
\end{aligned}
\end{equation} 
\begin{equation}
\setlength{\abovedisplayskip}{3pt}
\setlength{\belowdisplayskip}{3pt}
\begin{aligned}                                 
    \mathcal{L}_{joint}=\mathcal{L}_{item}+\gamma\mathcal{L}_{bhv}\,,
\end{aligned}
\end{equation} 
where $\mathbf{y} \in \mathbb{R}^{|V|}$ and $\mathbf{b} \in \mathbb{R}^{|R|}$ are one-hot vectors of ground truth for next item and next interaction type, and $\gamma$ is a control parameter for joint learning.
$\lambda_i$ indicates the loss weight of behavior $o_i$. 
For Task 1, we only optimize the item loss  $\mathcal{L}_{item}$ without the considering of next behavior prediction.  


\section{Experiment}

In this section, we  conduct extensive experiments on multi-behavior session-based recommendation to evaluate the performance of our model by answering the following four research questions\footnote{Our code and data will be released for research purpose.}:
\begin{itemize}
\item \textbf{RQ1:} Compared with other state-of-the-art MBSBR models, does our model achieve better performance?

\item \textbf{RQ2:} Is the introduction of multi-behavior sequence information efficient for the performance of our model?

\item \textbf{RQ3:} How do the key modules of MGCNet influence the model performance?

\item \textbf{RQ4:} How does the setting of hyper-parameters (such as the depth of GNN) affect the effectiveness of our model?

\end{itemize}

\subsection{Experimental Setup}
\subsubsection{Dataset}
We conduct extensive experiments on two benchmark datasets in the session-based recommendation research: \emph{Yoochoose} and \emph{Tmall-1 month} \cite{srgnn,beyondclick}. These datasets both contain the multiple types of user feedback (e.g. click and purchase) information that can support our work on multi-behavior session-based recommendation.
\begin{itemize}
    \item \emph{Yoochoose}\footnote{http://2015.recsyschallenge.com/challege.html} is derived from the RecSys Challenge 2015 and contains a stream of user actions on an e-commerce website over a period of six months.The operation types include click and buy.
    \item \emph{Tmall}\footnote{https://tianchi.aliyun.com/dataset/dataDetail?dataId=42} comes from IJCAI-15 competition, which contains anonymous user’s shopping logs on Tmall online shopping platform with click, add-to-favorite and buy interactions. We employ the data in May for our research.
\end{itemize}

Following \cite{liu2018stamp,narm}, we use the most recent fractions 1/64 and 1/4 of the training sequences of Yoochoose as the dataset \emph{Yoochoose 1/64} and \emph{Yoochoose 1/4}. For a fair comparison, following \cite{narm,srgnn}, we first filtered out all sessions of length $\leq2$ and items appearing less than 5 times in all datasets. Then we applied a data augmentation technique described in \cite{tan2016improved}.
For each dataset, we select about $20\%$ as the test set, $10\%$ as valid set and remaining $70\%$ as training set.
It is worth mentioning that there are no duplicate records for same items with different interaction types, i.e., the purchase behavior after the click interaction with same item is treated as purchase this item without the prior record of click behavior.
The statistics of all datasets after prepossessing are summarized in Table  \ref{tab:dataset}.
\begin{table}[htbp]
    \caption{Statistics of datasets used in experiments.}
    \label{tab:dataset}
     \small
    \centering
    \begin{tabular}
    {lrrr}
    \toprule
    Statistic& Yoochoose 1/64 & Yoochoose 1/4 & Tmall \\
    \midrule
    No. of items  &18,253  &  32,581 &43,908  \\
    No. of sessions  & 273,734  &3,967,318  & 881,262  \\
    Avg. of session length & 6.18 & 5.78 & 6.73 \\
    Ratio of click  &95.1\% & 94.9\%  & 90.3\% \\
    Ratio of purchase  &4.9 \% & 5.1\%  & 3.6\% \\
    Ratio of favorite & - & -  & 6.1\% \\
    
    \bottomrule
    \end{tabular}
\end{table}

\begin{table*}[]
\centering
    \caption{Experimental results (\%) of different models in terms of HR@20 and MRR@20 for different next behavior type session on three datasets. The * means the best results on baseline methods.}
    \label{tab:overall}
\begin{tabular}
{p{2.3cm}<{\centering}p{0.6cm}<{\centering}p{0.6cm}<{\centering}p{0.6cm}<{\centering}p{0.6cm}<{\centering}p{0.01cm}p{0.6cm}<{\centering}p{0.6cm}<{\centering}p{0.6cm}<{\centering}p{0.6cm}<{\centering}p{0.001cm}p{0.6cm}<{\centering}p{0.6cm}<{\centering}p{0.6cm}<{\centering}p{0.6cm}<{\centering}p{0.6cm}<{\centering}p{0.6cm}<{\centering}}
\hline
\multirow{3}{*}{\bfseries Models} & \multicolumn{4}{c}{\bfseries Yoochoose 1/64}& & \multicolumn{4}{c}{\bfseries Yoochoose 1/4}& & \multicolumn{6}{c}{\bfseries Tmall}                       \\ \cline{2-5} 
\cline{7-10} 
\cline{12-17} 
 &
  \multicolumn{2}{c}{Click} &
  \multicolumn{2}{c}{Purchase} & &
  \multicolumn{2}{c}{Click} &
  \multicolumn{2}{c}{Purchase} & &
  \multicolumn{2}{c}{Click} &
  \multicolumn{2}{c}{Purchase} &
  \multicolumn{2}{c}{Favorite} 
  \\ 
 \cline{2-5} 
\cline{7-10} 
\cline{12-17} 
& HR  & MRR  & HR  & MRR & & HR & MRR  & HR  & MRR &  & HR & MRR & HR & MRR & HR & MRR \\ 
\hline\hline
GRU4Rec & 61.42 & 29.87  & 70.42  & 39.65&  & 60.93 & 29.69  & 70.73 & 29.31& &19.74  &8.27  & 54.24  & 29.32  & 32.47 &14.02        \\ 
NARM  & 69.21 & 33.28   &81.01  &48.97&  &69.38  &34.11   &81.46  &49.59 & &30.80  &13.68  &56.16  &31.73  &33.18     &15.32        \\ 
SRGNN  &70.52  &37.21   &89.19  &70.44&  &70.89  &37.43  &89.30  &71.51 & & 27.57 &14.34  &57.78   &33.69  & 33.58 & 16.02       \\ 
LESSR  &71.02*  &38.16*   &91.21*  &74.42*&  &71.43*  &38.33*   &91.94*  &74.36*&  &31.62*  &14.52*  &58.24*  &34.63* & 34.49* &16.78*        \\  \midrule
MGNN-SPred &56.94  &27.20   &81.58  &51.67&  &57.58  &27.48   &81.95  &51.69&  &29.42  &13.27  &55.82 &32.46  &33.86  &15.39        \\ 
MBGCN  & 57.85 &28.49   & 80.07  &51.21&  &56.92  &26.87   &80.13  &51.14 & &26.12   &12.33  &53.27   & 31.19 & 30.95 & 13.63        \\
MKM-SR  &70.71  &37.82  &89.47  &71.07&  &71.31  &37.82   &89.41 &71.42&  &27.18  & 14.11  & 57.90 & 33.94  & 34.42  &16.75         \\  
 \midrule
 Ours &{\bfseries 72.28} &{\bfseries 38.46} &{\bfseries 92.38} &{\bfseries 74.90}& &{\bfseries 72.40} &{\bfseries 39.39} &{\bfseries 92.96} &{\bfseries 76.37}& &{\bfseries 32.63} &{\bfseries 15.09} &{\bfseries 63.15}  &{\bfseries 40.28} &{\bfseries 37.12} &{\bfseries 18.46}    
\\ 
 \midrule
 Ours + GT  &72.67  &38.97  &93.27  &75.59&  &72.51  &40.52   &93.36 &77.05 &  &32.91&15.51&63.42&40.51&37.53&18.81 
\\  \bottomrule
\end{tabular}
\end{table*}

\subsubsection{Baseline Models}
To demonstrate MGCNet’s superiority performance, we compared it with the  following representative methods for MBSBR:
\begin{itemize}
    \item \textbf{GRU4Rec} \cite{gru4rec} utilized the GRU to capture the representation of the item sequence.
    \item \textbf{NARM} \cite{narm} adopted RNN to encode the item representation and an attention mechanism  based on last item to capture the sequential feature.
	\item \textbf{SRGNN} \cite{srgnn} utilized GGNN to capture the item transition patterns and generated the session representation based on the attention module.
	\item \textbf{LESSR} \cite{lessr} proposed to learn the item embeddings from an edge-order preserving graph and a shortcut graph iteratively based on current session.
	\item \textbf{MGNN-SPred} \cite{beyondclick} converted sequences into a global item-item graph and exploited the GNNs to learn item representations, and integrated different behavior session representations by the gating mechanism.
	\item \textbf{MBGCN} \cite{MBGCN} constructed a heterogeneous graph with user-item and item-item relations to model  multi-behavior interactions.
	\item \textbf{MKM-SR} \cite{MKMSR} employed GGNN to encode the item transitions and RNN to encode the behavior feature separately. A session representation was learnt by the attention mechanism based on the combination of the above two features.

\end{itemize}

For original GRU4Rec, NARM, SRGNN, and LESSR methods, they are developed for the common session-based recommendation, which only consider single click behavior.
And multi-behavior based models like MGNN-SPred and MKM-SR, concentrate on next item prediction with specific target behavior based on historical multiplex interaction sequence.
To make the comparison fairer, we revise these methods in the following manner.
We use the original forms to model the multi-behavior sequence with the input of item and behavior type simultaneously for RNN based models. 
For GNN based models, we additionally concatenate corresponding current behavior embedding with the item representation after GNN encoders.
Here, we only employ the item-item GNN part of MBGCN and remove the user-related module following the definition of SBR, and reform the item-item edges with same behavior within one session.
Moreover, we take all multiplex next behavior data for unified training.

\subsubsection{Implementation Details}
We implement the proposed model based on Pytorch and DGL.
The item and behavior embedding dimension is set to 128 for all models.
All parameters are initialized using a Gaussian distribution with a mean of 0 and a standard deviation of 0.1.
We use the Adam optimizer to train the models with the mini-batch size of 512. We conduct the grid search over hyper-parameters as follows: learning rate $\eta$ in $\{0.001,0.01,0.1\}$, learning rate decay in $\{0.01,0.05,0.1,0.5\}$, learning rate decay step  in $\{2,3,4\}$.
The hyper-parameter $\gamma$ is set to $10$ in $\{0.1,1,,10,100\}$.
About the hyper-parameters $\lambda$ in loss function, we set $\{\lambda _{click},\lambda_{purcahse},\lambda_{fav}\}=\{0.2,0.4,0.4\}$ for Tmall dataset and $\{\lambda _{click},\lambda_{purcahse}\}=\{0.2,0.8\}$ for two Yoochoose datasets based on the distributions of different next behavior types,  separately.
For the other baselines, in order to achieve the best performance for MBSBR, we tune the hyperparameters of these revised methods, with the same ranges of our tuning experiments.

\subsubsection{Evaluation Metrics}
We evaluate the performance of all models by employing the following widely used metrics: 
\begin{itemize}
    \item Hit Ratio (HR@K) is widely used as a measure of predictive accuracy. It indicates the proportion of correctly recommended items among the top-$K$ items.
    \item Mean Reciprocal Rank (MRR@K) is the average of the reciprocal ranks of the 
target items. The reciprocal rank is set to zero if the rank exceeds $K$.
\end{itemize}

For multiple next behavior data, we separately calculate the metrics with different next behavior type sessions, while train a uniform recommendation model for all kinds of next behavior.





\begin{table}[htbp]

    \caption{The recall result(\%) for various next behavior prediction of Task 2. }
    \label{tab:recall}
    \centering
     \small
    \setlength{\tabcolsep}{2pt}
    \begin{tabular}{lrrrrrrrrrr}
    \toprule
   
\multirow{2}{*}{\bfseries Dataset} & &
  \multicolumn{2}{c}{Yoochoose 1/64} & &
  \multicolumn{2}{c}{Yoochoose 1/4}  & &
  \multicolumn{3}{c}{Tmall}  \\ 
  \cline{3-4} \cline{6-7} \cline{9-11} 
   & & Click & Purchase & & Click & Purchase& & Click & Purchase &Favorite  \\
    \midrule
 Recall & &95.5 &68.1 & &96.2	&68.4 & &93.1&	72.7		&69.8
\\
    \bottomrule
    \end{tabular}
\end{table}

\subsection{Overall Comparison}

The comparison results on three datasets are reported on \autoref{tab:overall}. 
"Ours + GT" indicates our model for Task 1 with the given next behavior.
We can obtain the following significant observations.
\begin{itemize}
\item \textbf{Model Effectiveness.} From above table, we can find that MGCNet comprehensively outperforms all baselines on all metrics for different behaviors.
The results indicate that our model is effective to the MBSBR.



\item \textbf{Our model can leverage the global graph for session recommendation efficiently.}
At the multi-behavior category, MGCNet outperforms global graph based MGNN-SPred and adapted MBGCN in all metrics with different next behavior types. 
This demonstrates the strong power of behavioral contextual aware item representation layer, since global intention and local interest are integrated into final item representation, which empowers the effectiveness of representation learning for MBSBR.
However, MGNN-SPred and adapted MBGCN fail to leverage local intention signals for item representation. 

\item\textbf{Our model achieves competitive performance among the multi-behavior based approaches.}
Comparing with the multi-behavior based models such as  MGNN-SPred, MKM-SR, our MGCNet performs the best. 
For MGNN-SPred, it leverages the multiplex behavior data simplistically, which may lead to the information loss in original data.
For MKM-SR, it only encodes the behavior sequence independently but ignores the contribution of corresponding items. 
Meanwhile, all the above methods ignore the potential relation between next behavior and historical operation sessions.
In contrast, MGCNet can extract intention strength and semantic relation of interest based on former interaction sequence and next behavior intention, which strongly improves the model effectiveness.
\end{itemize}

Moreover, \autoref{tab:recall} represents the next behavior prediction result for Task 2, measured with recall metrics on various target behavior sessions.
Compared with the results of "Ours" and "Ours + GT", we find that the ground truth behavior information offered in Task 1 can offer more accurate intention information than the forecasted behavior tag, which suggests that a better grasp of the  next behavior will make better guidance for the next item prediction.

\subsection{Ablation Study (RQ2 \& RQ3)}

In this subsection, we conduct some ablation studies on the proposed model to investigate the effectiveness of behavior information and some module designs. 

\begin{table}[htbp]
    \caption{The performance comparison w.r.t different information on \emph{Tmall}. }
    \label{tab:information}
    \centering
    \setlength{\tabcolsep}{2pt}
    \begin{tabular}{lrrrrrr}
    \toprule
   
\multirow{2}{*}{\bfseries Model setting} &
  \multicolumn{2}{c}{Click} &
  \multicolumn{2}{c}{Purchase}  &
  \multicolumn{2}{c}{Fav}  \\ \cline{2-7} 
    & HR &MRR &HR &MRR &HR &MRR  \\
    \midrule
    {NARM-Single}&30.62 &13.41 &55.72 &31.29 &32.73 &15.04\\
    {NARM-Multi}&30.80 &13.68 &56.16 &31.73 &33.18 &15.32\\   
    {LESSR-Single}&31.49 &14.47 &58.22 &34.57 &34.44 &16.82\\
    {LESSR-Multi}&31.64 &14.52 &58.24 &34.63 &34.49 &16.78\\   
    \midrule
    Ours-Single &31.52 &14.38 &59.47 &35.94 &35.16 &16.25\\
    Ours w/o Next &32.17	&14.72	&60.29	&37.83	&35.62	&16.70 \\
    Ours &32.63	&15.09	&63.15	&40.28	&37.12	&18.46\\
    Ours + GT &32.91&15.21&63.32&40.35&37.43&18.61 \\
    \bottomrule
    \end{tabular}
\end{table}

\subsubsection{Impact of behavior information}
We compare our model with several representative baselines in \autoref{tab:information} to test whether considering multiplex behavior sequence indeed boosts the performance of SBR.
The methods with “(-Single)" mean considering multiplex interaction types as single type from their full version, while "(-Multi)" is the multi-behavior based version.
By comparing each method in  \autoref{tab:information} with two versions, we find each method with multi-behavior information beats single behavior based model in most metrics on \emph{Tmall} dataset. 
Based on the above illustrations, we demonstrate that considering the multiplex behavior interaction sequence is indeed meaningful.

Moreover, we conduct ablation studies of our model with various next behavior information. 
"Ours w/o Next" indicates the loss the next interaction type prediction in our model.
The advantage of "Ours + GT" and "Ours" over "Ours w/o Next" shows that the next behavior type deserves to be incorporated for SBR.

\begin{table}[htbp]
    \caption{The performance comparison w.r.t. different module design on \emph{Tmall}. }
    \label{tab:module}
    \centering
    \setlength{\tabcolsep}{2pt}
    \begin{tabular}{lrrrrrr}
    \toprule
   
\multirow{2}{*}{\bfseries Model setting} &
  \multicolumn{2}{c}{Click} &
  \multicolumn{2}{c}{Purchase}  &
  \multicolumn{2}{c}{Fav}  \\ \cline{2-7} 
    & HR &MRR &HR &MRR &HR &MRR  \\
    \midrule 
    {w/o Global Graph}&31.62 &14.03 &62.06 &39.22 &36.30 &17.76\\
    
    {w/o BCIRL } &31.72 &14.31 &62.41 &39.53  &36.40  &17.93  \\
    
    {w/o Current attention} &32.14 &14.73 &60.36 &36.95 &36.41  &17.27  \\
    {w/o General attention}&31.67 &14.29 &61.90 &38.71 &36.84 &17.77\\
 
    \midrule
     MGCNet&32.63 & 15.09 & 63.15 & 40.28 &37.12 &18.46 \\
    \bottomrule
    \end{tabular}
    
\end{table}

\subsubsection{Impact of different Layers}
In this part, we compare our method with different variants to verify the effectiveness of the critical components of MGCNet. 
Specifically, we remove critical modules of MGCNet to observe changes in model performance, using "w/o Global Graph" to denote removing the global heterogeneous graph encoder, and using "w/o BCIRL" to denote skipping the Behavioral contextual aware item representation learning layer.
The experimental results are presented in \autoref{tab:module}. 
It can be observed that the item representations of the global graph are pivotal for the model performance by seeing  "w/o Global Graph".
For the session intention modeling, the removal of general attention module leads to a greater impact on results than the current attention module, which demonstrates that long-term user preference is still valuable for SBR. 
In summary, we can infer that the key components of MGCNet are effective through the above comparison and analysis.

\begin{figure}[t]
    \centering
    \begin{subfigure}{0.48\linewidth}
        \includegraphics[width=\textwidth]{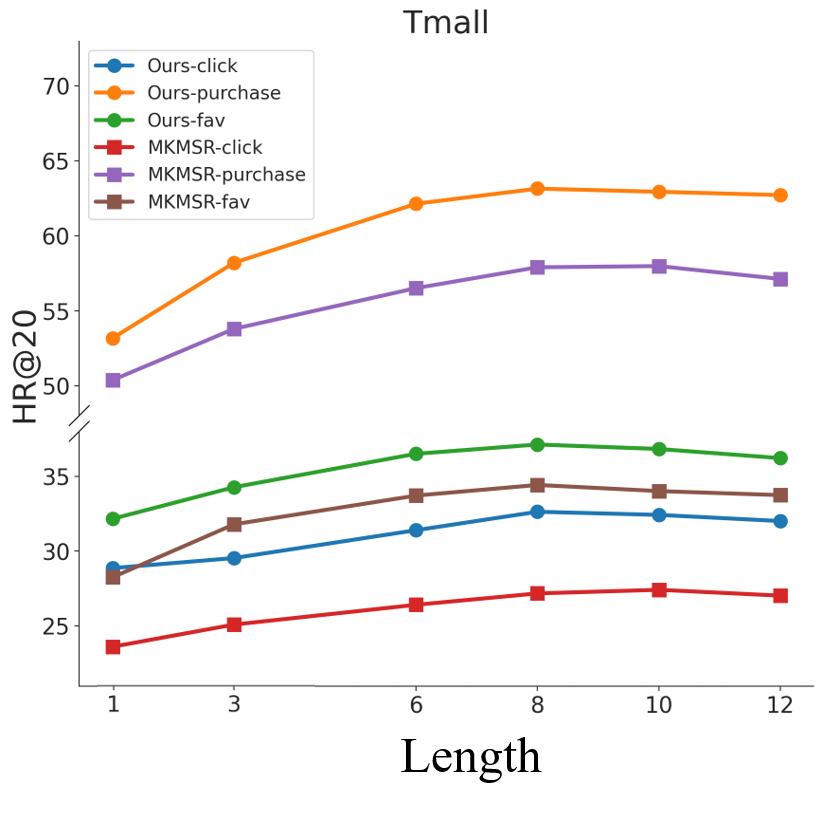}
    \end{subfigure}
    \begin{subfigure}{0.48\linewidth}
        \includegraphics[width=\textwidth]{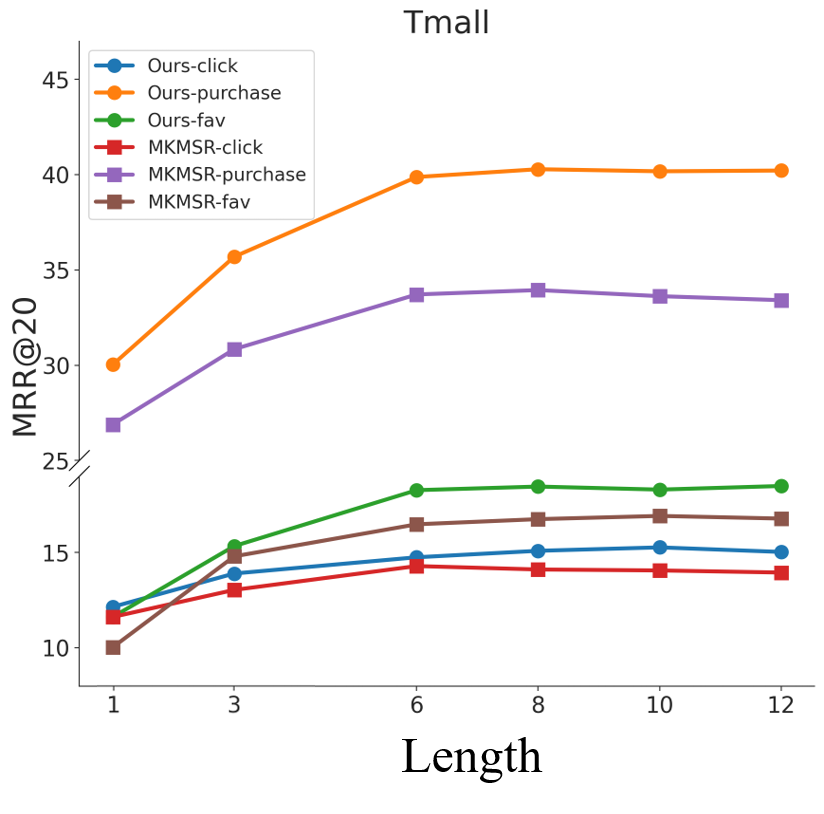}
    \end{subfigure}
    \caption{Performance comparison w.r.t. the max session length.}
    \label{fig:hyper-length}
\end{figure}

\begin{figure}[t]
    \centering
    \begin{subfigure}{0.48\linewidth}
        \includegraphics[width=\textwidth]{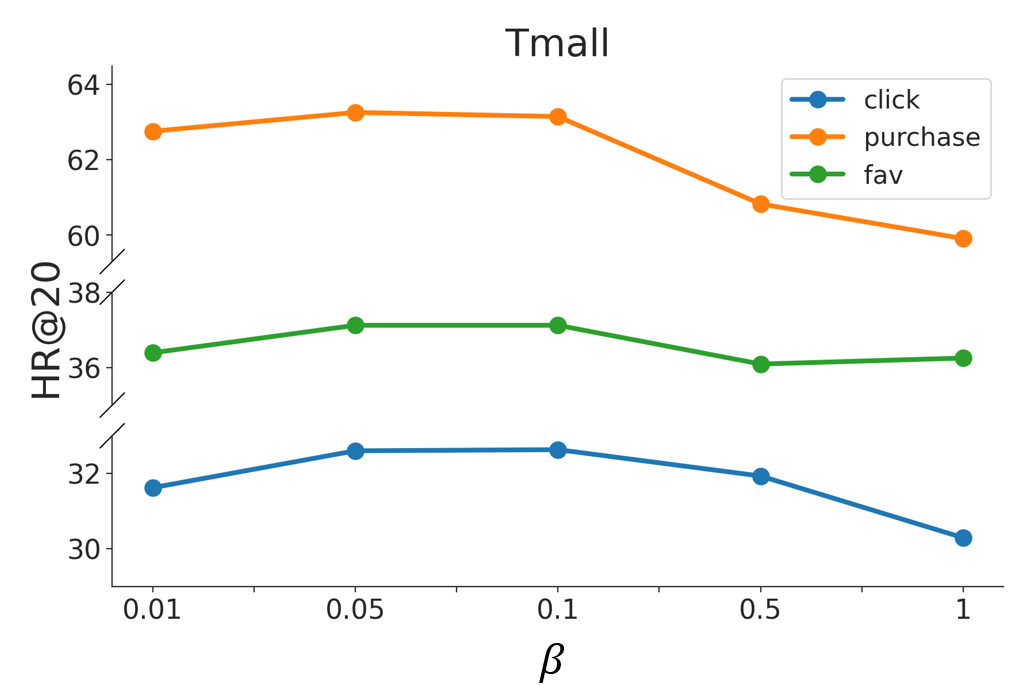}
    \end{subfigure}
    \begin{subfigure}{0.48\linewidth}
        \includegraphics[width=\textwidth]{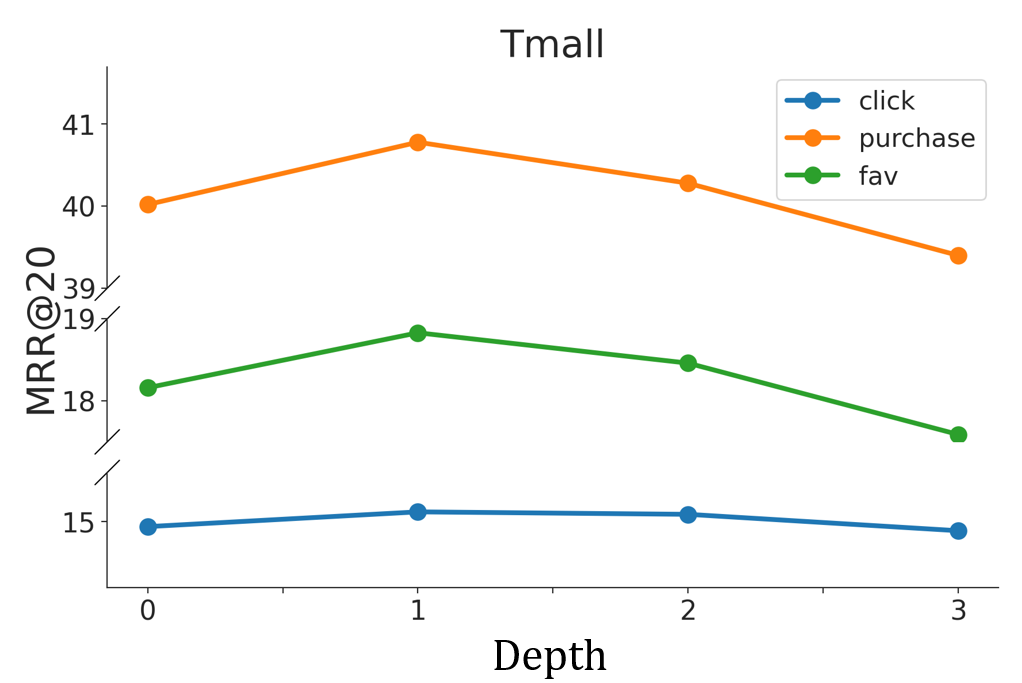}
    \end{subfigure}
    \caption{Performance comparison w.r.t. $\beta$ and GNN layers.}
    \label{fig:hyper-beta}
\end{figure}

\subsection{Hyper-parameters Study (RQ4) }\label{sec:hyper-res}

In this subsection, we perform experiments to explore how the hyper-parameters like GNN layers and intention factor $\beta$ influence the model performance.

\subsubsection{Impact of maximum session length}

In this part, we discuss how the performance changes with hyper-parameter maximum session length $M$, which indicates the upper limits of historical information that the network can utilize to make predictions for current session. 
\autoref{fig:hyper-length} shows the evaluation results with different maximum session lengths in the range from $1$ to $12$. 
HR@20 reaches the highest score when $M$ is 8 for \emph{Tmall}. 
Overall, our model outperforms MKM-SR consistently.
As expected, with larger maximum sequence length at the beginning, the performance of both our model and MKM-SR grows to be better. 
That show effectiveness of sequence information for learning current user interest.
In general, we can find that longer session does not lead to better performance, which indicates that the increase of current session length does not necessarily lead to an increase in model performance.

\subsubsection{Impact of Depth of GNN}
We test different depth settings about graph representation propagation. 
The depth setting with value 0 indicates that our model does not use multi-relation GNN and only learns the item representation based on the naive item embedding.
Figure \ref{fig:hyper-beta} shows the corresponding results. 
We can see that the performance of depth 0 is worse than multi-layers of GNN.
This comparison clarifies the significance of considering the global item and behavior transitions for our model.
Moreover, the performance declines when the depth grows from 1 to 3, showing the overly depth of GNN layers will lead to over-smooth problem and make the item representation less distinguishable, which is not ideal for further improving the performance.

\subsubsection{Impact of Intention Factor} 
We investigate the influence of relevant local intention to item final by intention factor $\beta$ for session recommendation at Item Representation Learning module. Fig. \ref{fig:hyper-beta} visualizes  MGCNet's HR@20 scores in Tmall, from which we find that MGCNet's performance varies marginally ($\sim 1\%$) when $\beta$ is set in $\left[0.05, 0.1\right]$. What’s more, the performance of MGCNet achieves the peak when $\beta=0.1$, and sharply drops when $\beta\geq0.1$, which implies that overweight assigned with local intention may disturb the session representation for MBSBR.

\section{Conclusion}
In this work, due to the limitations of the existing tasks, we propose  two  more  realistic  and  reasonable  tasks  for MBSBR.
To address the above tasks, we develop a Multi-behavior Graph Contextual Aware Network (MGCNet), which explicitly utilizes the next behavior and integrates item representation of global transitions overall sessions and local intention  based on current session for better session recommendation. 
Extensive experiments and ablation studies on three datasets demonstrate the superior performance of our model.

In the future, we plan to explore the fine-grained intent of different interaction behaviors within the current session. 
We also plan to introduce more side information, such as time, brand information of items, etc.

\bibliographystyle{ACM-Reference-Format}
\bibliography{ref}



\end{document}